\documentstyle[11pt,newpasp,twoside,epsf]{article}
\def\edcomment#1{\iffalse\marginpar{\raggedright\sl#1\/}\else\relax\fi}
\reversemarginpar

\begin{document}
\title{The Environmental Dependence of Star Formation in the Nearby Universe}
\author{Laerte Sodr\'e Jr.}
\affil{Departamento de Astronomia, IAG, Universidade de S\~ao Paulo, Brazil}
 \author{Ab\'\i lio Mateus Jr.}
\affil{Departamento de Astronomia, IAG, Universidade de S\~ao Paulo, Brazil}

\begin{abstract}
We have investigated how the incidence of star-forming galaxies varies with
the environment using 
a volume-limited sample of nearby galaxy spectra extracted from the 
2dF Galaxy Redshift Survey. The environment is characterized by the local 
number density of galaxies, corrected for sample incompleteness and
border effects. Using the equivalent widths of [O II]$\lambda$3727
and H$\delta$ we discriminate the star-forming galaxies in 
two classes, starbursts and ordinary star-forming galaxies,
and evaluate their properties as a function of the local density. 
We show that the fraction of ordinary star-forming galaxies decreases
regularly with increasing density, indicating that star formation is 
sensitive to the environment for all range of densities present in the sample.
The fraction of starbursts is approximately 
independent of the
density and, consequently, the fraction of starbursts
relative to star-forming galaxies increases with  local density. 
This suggests that a mechanism that acts everywhere, like tidal
interactions, is the responsible for triggering starbursts.
We also show that while the mean EW(H$\alpha$) of  ordinary star-forming 
galaxies is progressively reduced as density increases, for starbursts it
suffers a strong decrease at densities corresponding to scales of 
$\rho^{-1/3} \sim 2.5 ~h^{-1}$ Mpc. A
visual inspection of starburst images in the {\it Digital Sky Survey} 
reveals that this
corresponds to a change in the morphology of starburst galaxies, 
from mainly apparently normal spirals and galaxies with apparently
compact morphologies at low densities,
to tidally distorted spirals at high densities.

\end{abstract}

\section{Introduction}

Studies of the effects of the large scale environment around a galaxy on the
properties of its
star formation are essential for understanding how galaxies evolve.
It is well known since long time that galaxies with large star formation
rates avoid dense environments, and many recent works
have confirmed that star formation is lower for galaxies in clusters and
groups than for
those in the field (e.g., Ellingson et al. 2001, Mart\' \i nez et al. 2002).
Several different process have been proposed to explain why and how star
formation is affected by the environment, like mergers, ram pressure, 
and tidal 
interactions. While all of them may take place in some form or another, the 
main process that affects star formation in galaxies is not clear yet.

In the current era of large redshift surveys, enormous amounts of homogeneous
data on nearby 
and distant galaxies are becoming available and may be used to study
statistically many 
aspects of star formation in galaxies. Here we use the first 100k Data Release
of the 2dF Galaxy Redshift Survey (2dFGRS; Colless et al. 2001) to examine
some properties of star-forming  galaxies in the nearby universe. 

\section{A Nearby Galaxy Sample}
The sample analyzed here was extracted from the 2dFGRS. We adopt
a volume limited sample because, in this case, the selection function is 
in principle uniform and the variations in galaxy number density are due 
to galaxy clustering only. 
We have considered  galaxies with velocities between
600 km s$^{-1}$ and 15000 km s$^{-1}$ brighter than the extinction
corrected magnitude $b_J=18.50$ (corresponding to 
$M_{b_J}^{lim}=-17.38+5\log h$) 
and located within the SGP and NGP strips covered by the survey. 
Galaxy distances were computed from the redshifts, neglecting 
peculiar velocities and $k$-corrections.

The redshift completeness and
its magnitude dependence vary over the area of the survey but,
nevertheless, they can be represented by a mask 
that gives the selection function at each position in the survey area
(Colless et al. 2001, Norberg et al. 2001). We have used numerical 
simulations to study how the survey geometry and its current incompleteness 
affect our 
estimates of the galaxy density (Mateus \& Sodr\'e, in preparation).
These simulations allowed us to estimate the mean local correction factor, 
${\cal C}$, that should be applied to the observed density to correct
it of local incompleteness and border effects (${\cal C} \ge 1$).
To minimize these corrections and, 
at the same time, to have a sample as large as possible, we have considered
only regions for which ${\cal C} \le 2.5$.

\section{Determination of the Local Galaxy Density}
We characterize the environment of a galaxy by the three-dimensional
galaxy number density in its neighborhood. This density is calculated
from the distance $r$ to its $5-$th nearest 
neighbor  in the sample and corrected
with the local value of ${\cal C}$. Since we are neglecting
peculiar velocities, galaxies in dense environments, where the velocity
dispersion is high, will probably have their local densities underestimated.
However, only one of the cataloged galaxy clusters in the survey region 
(De Propis et al. 2002), S0006, is inside our selected volume. Hence,
our sample is more representative of the field than of the cluster
galaxy population.

The local densities (of galaxies brighter than $M_{b_J}^{lim}$) 
were computed for all the galaxies in the sample above.
It can be shown that our results are not affected, qualitatively, by the
density corrections.

\section{Spectral Classes}
Spectra obtained in modern redshift surveys like the 2dFGRS have much more
information than that needed to obtain just redshifts, and can be used to 
provide informations on the galaxy type or star formation activity.
The use of equivalent widths (EWs) 
of certain spectral features are particularly useful in the study of 
survey data because they are not affected by reddening and are
relatively insensitive to the instrumental resolution. 

Following Balogh et al. (1999), we use the EWs of the [O II]$\lambda$3727 
doublet and of $H\delta$ to classify galaxies in distinct spectral classes.
The EW of [O II] measures the activity of star formation, whereas the
EW($H\delta$) is sensitive to the age of star formation bursts. 
EWs in the rest-frame of the objects were computed and 146 of them that
presented bad quality spectra or were classified as AGN (i.e, with
EW($H\alpha$) $>$ 10 \AA~ and  EW([N II]) $>$ 0.55 EW($H\alpha$); Lewis
et al. 2002) were removed from the analysis. The remaining sample, with
$N_{ALL}=1234$ galaxies, will be analyzed hereafter.

We consider star-forming galaxies those with EW([O II]) larger than 5 \AA,
roughly the detection limit of this spectral feature in the survey spectra. 
Notice that we adopt here positive and negative 
values for the EW of emission and absorption
lines, respectively. The spectral classes defined in the 
plane EW($H\delta$) vs. EW([O II]) and discussed in this work are:
\begin{itemize}
\item non star-forming galaxies (P: EW([O II]) $\le$ 5 \AA) \\
passive galaxies, without evidence of current significant star formation
(in general E or S0);
\item starburst galaxies (SB: EW([O II]) $>$ 5\AA, 
EW($H\delta$) $>$ 0) \\
galaxies where  a large fraction of the light comes from a starburst
that started less than $\sim$200 Myr ago; these galaxies are called SSB
(short starburst) by Balogh et al. (1999);
\item ordinary star-forming galaxies (SF: EW([O II])$>$5\AA, 
EW($H\delta$)$<$0) \\
includes most normal spirals and irregulars, that have been forming stars 
for several hundred million years.
\end{itemize}
With these definitions, the total number of galaxies 
with star formation is $N_{TSF}=635$, of which $N_{SB}=66$ and $N_{SF}=569$, 
and of those without evidence of ongoing star formation is $N_{P}=599$.
Considering only galaxies with star formation, SBs correspond to  10.4\% 
of the total.

\section{Type vs. Local Density}
Our main results are summarized in Figure 1. In all plots the data of 
each class was divided in six density bins with approximately the same number 
of galaxies per bin. Fig. 1a shows that the fraction of star-forming galaxies 
decreases steadily with increasing local density. 
At the same time, the fraction of non star-forming galaxies
increases with increasing density. These trends
may be seen as a consequence of the well-known density - morphology relation 
proposed by Dressler (1980). 
The fraction of SBs, however, is essentially independent of local
density. Since the fraction of star-forming galaxies decrease with
increasing density, in dense environments SBs become relatively more frequent 
among star-forming galaxies.
 
The relation between the mean value of EW(H$\alpha$) and local density,
shown in Fig. 1b, also provides interesting 
information about the behavior of star formation in distinct environments. 
Passive galaxies do not have significant H$\alpha$ emission, as expected.
For ordinary star-forming galaxies the EW(H$\alpha$)  is essentially constant
with density. For the SBs, however,
the EW(H$\alpha$) presents a clear decrease at $\rho \sim 0.06~h^3$ 
Mpc$^{-3}$. At lower densities these galaxies 
have mean EW(H$\alpha$) larger than those 
of SF galaxies. For higher densities, however,
it drops from $\sim$ 30 \AA~ to $\sim$ 10 \AA, below the mean value of SFs,
$\sim$ 20 \AA.

\section{Discussion}

In the hierarchical scenario, as galaxy clustering evolves, the density 
around a galaxy tends to increase, in all environments. Higher density
probably means more interactions. Our results give support to this broad view
because star formation seems to be affected by the environment in all range
of densities covered by our sample, as evidenced by the decrease in
the fraction of star-forming galaxies with increasing local density
shown in Fig. 1. Lewis et al. (2002), analyzing a different sample 
extracted from the 2dFGRS,  find
a density threshold, below which the star formation is insensitive to the
galaxy environment. This is in variance with our results, which suggest that
the processes that affect the star formation activity in a galaxy do act 
everywhere, although they may be different in distinct environments.

Most of star-forming galaxies have their star formation activity inhibited 
or even interrupted as density increases, but the fraction of SBs seems to
be roughly independent of local density, contrarily to the expectation that
denser environments favor the triggering of a starburst. However, since the 
fraction of star-forming galaxies decreases in dense regions, 
the fraction of SB galaxies relative to all star-forming galaxies is 
actually enhanced. 
This behavior have been noticed already in a sample of galaxies in the 
Shapley supercluster (Cuevas 2000). 
From a study of 8 Abell clusters, Moss \& Whittle (2000) have also concluded 
that the fraction of SBs among spirals increases from regions of lower to 
higher local
galaxy surface density. They noticed that in most starbursts the 
H$\alpha$ emission is circumnuclear and the galaxy morphology is disturbed,
indicating the action of tidal interactions;
these galaxies are the local counterparts of the Butcher-Oemler objects.
Here we have shown that the fraction of SBs is essentially the same in all 
environments. This favors mechanisms like tidal interactions for triggering
starbursts because they act everywhere, and not only in high-density
environments, like ram-pressure stripping or high-speed interactions
between galaxies (`harassment').

We have also found that the EW(H$\alpha$) of starbursts suffers a
strong reduction in high density regions. This reduction occurs at a 
characteristic
intergalactic distance ($\rho^{-1/3}$) of $ \sim 2.5 ~h^{-1}$ Mpc. This is
a quite large separation, when compared with the distances between the 
members of our Local Group. 

We have inspected visually the sample of starburst galaxies using 
{\it Digital Sky Survey} images. A clear trend with morphology is present: 
at low densities most of the starbursts are apparently normal
spirals or galaxies with an apparently compact morphology, whereas at 
high densities the starburst class contains a large fraction of tidally 
distorted spirals. In the lowest density bin only 1 out of 11 galaxies is
clearly a tidally distorted spiral; in the highest density bin, however,
6 out of 11 are galaxies of this kind. This suggests that tidal interactions
in dense environments may indeed favor the triggering of starbursts in 
spiral galaxies.

\begin{figure}
\plotone{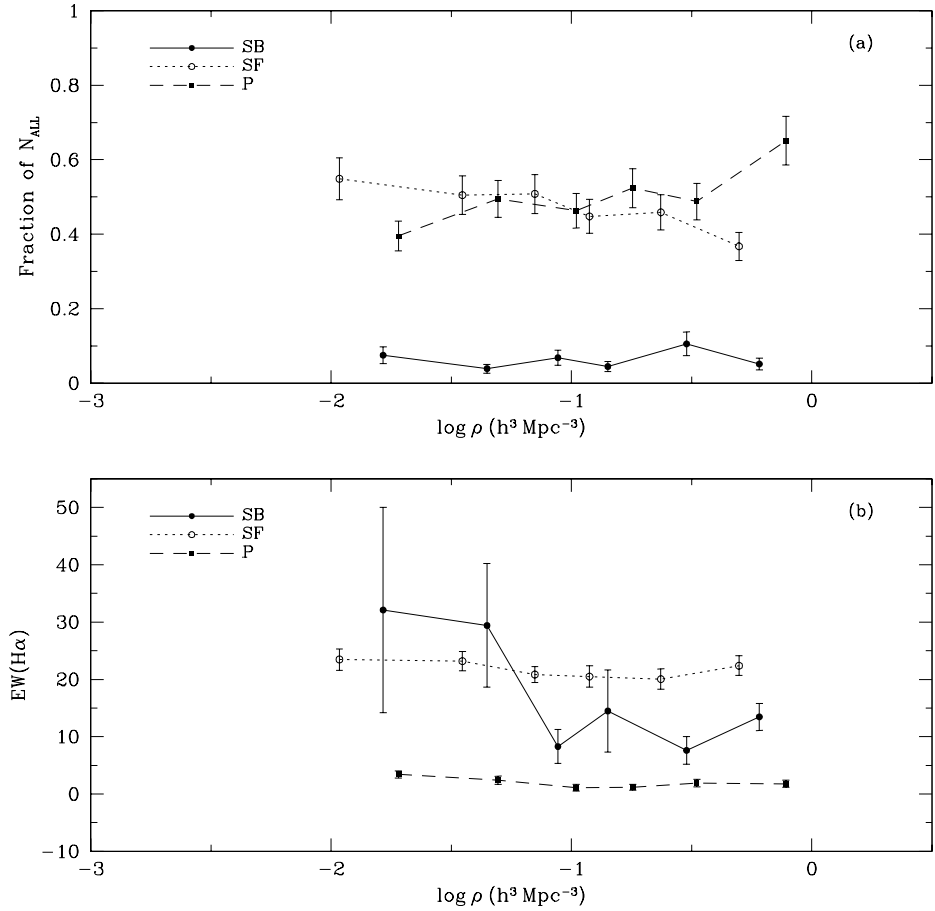}
\caption{Some properties of the sample as a function of the local galaxy
number density. For each class, the density was divided in 6 bins with
approximately the same number of galaxies per bin. 
a) Fractions of galaxy classes; the error bars assume Poisson statistics.
b) EW(H$\alpha$), with the errors of the mean.}
\end{figure}

\acknowledgements
We thank FAPESP, CNPq, PRONEX, and CCINT/ USP for supporting this work.

\end{document}